\documentclass[reprint,amsmath,amssymb,aps,nofootinbib]{revtex4-2}
\usepackage{amssymb}
\usepackage{subfigure}
\usepackage{xspace}
\usepackage{textpos}
\usepackage[english]{babel}
\usepackage{graphicx}
\usepackage{dcolumn}
\usepackage{bm}
\usepackage[colorlinks,linkcolor=blue,anchorcolor=blue,citecolor=blue,urlcolor=blue]{hyperref}
\usepackage{overpic}
\usepackage{multirow}
\usepackage{color}
\usepackage{cancel}
\usepackage{ulem}
\usepackage{amsmath}
\usepackage{makecell}
\usepackage{booktabs}
\usepackage{array}

\newcommand{\BESIIIorcid}[1]{\href{https://orcid.org/#1}{\hspace*{0.1em}\raisebox{-0.45ex}{\includegraphics[width=1em]{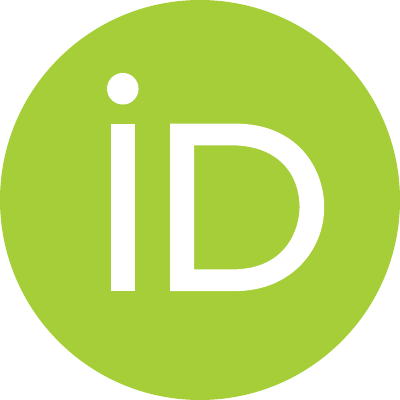}}}}

\begin{document}

\title{Measurement of the Absolute Branching Fraction of \textbf{\boldmath $\Xi(1530)^{-}\to (\Xi\pi)^{-}$} and Updated Measurement of the Branching Fraction of \textbf{\boldmath $\psi(3686)\to\bar{\Xi}^{+}\Xi(1530)^{-}+\mathrm{c.c.}$}}

\author{
\begin{small}
\begin{center}
M.~Ablikim$^{1}$\BESIIIorcid{0000-0002-3935-619X},
M.~N.~Achasov$^{4,c}$\BESIIIorcid{0000-0002-9400-8622},
P.~Adlarson$^{81}$\BESIIIorcid{0000-0001-6280-3851},
X.~C.~Ai$^{87}$\BESIIIorcid{0000-0003-3856-2415},
C.~S.~Akondi$^{31A,31B}$\BESIIIorcid{0000-0001-6303-5217},
R.~Aliberti$^{39}$\BESIIIorcid{0000-0003-3500-4012},
A.~Amoroso$^{80A,80C}$\BESIIIorcid{0000-0002-3095-8610},
Q.~An$^{77,64,\dagger}$,
Y.~H.~An$^{87}$\BESIIIorcid{0009-0008-3419-0849},
Y.~Bai$^{62}$\BESIIIorcid{0000-0001-6593-5665}
% ...
\\
\vspace{0.2cm}
(BESIII Collaboration)\\
\vspace{0.2cm} {\it
$^{1}$ Institute of High Energy Physics, Beijing 100049, People's Republic of China\\
% ...
$^{4}$ Budker Institute of Nuclear Physics SB RAS (BINP), Novosibirsk 630090, Russia\\
% ...
$^{31}$ INFN Sezione di Ferrara, (A)INFN Sezione di Ferrara, I-44122, Ferrara, Italy; (B)University of Ferrara, I-44122, Ferrara, Italy\\
% ...
$^{39}$ Johannes Gutenberg University of Mainz, Johann-Joachim-Becher-Weg 45, D-55099 Mainz, Germany\\
% ...
$^{62}$ Southeast University, Nanjing 211100, People's Republic of China\\
% ...
$^{64}$ State Key Laboratory of Particle Detection and Electronics, Beijing 100049, Hefei 230026, People's Republic of China\\
% ...
$^{77}$ University of Science and Technology of China, Hefei 230026, People's Republic of China\\
% ...
$^{80}$ University of Turin and INFN, (A)University of Turin, I-10125, Turin, Italy; (B)University of Eastern Piedmont, I-15121, Alessandria, Italy; (C)INFN, I-10125, Turin, Italy\\
$^{81}$ Uppsala University, Box 516, SE-75120 Uppsala, Sweden\\
% ...
$^{87}$ Zhengzhou University, Zhengzhou 450001, People's Republic of China\\
\vspace{0.2cm}
$^{\dagger}$ Deceased\\
% ...
$^{c}$ Also at the Novosibirsk State University, Novosibirsk, 630090, Russia\\
% ...
}
\end{center}
\vspace{0.4cm}    
\end{small}
}

\begin{abstract}
Based on $(2712.4\pm14.3)\times10^{6}$ $\psi(3686)$ events collected with the BESIII detector, the decays $\Xi(1530)^{-}\to\Xi^{0}\pi^{-}$ and $\Xi(1530)^{-}\to\Xi^{-}\pi^{0}$ are investigated jointly via the process $\psi(3686)\to\bar{\Xi}^{+}\Xi(1530)^{-}+\mathrm{c.c}$. Under the assumption of isospin symmetry, the two decay modes are treated as fully correlated, and we report the first measurement of their absolute branching fractions. The results are $\mathcal{B}(\Xi(1530)^{-}\to\Xi^{0}\pi^{-})=(61.4\pm4.5\pm4.6)\%$ and $\mathcal{B}(\Xi(1530)^{-} \to \Xi^{-}\pi^{0})
=(29.7\pm2.2\pm2.2)\%$. The combined branching fraction of the two decays is $\mathcal{B}(\Xi(1530)^{-}\to(\Xi\pi)^{-})=(91.1\pm6.7\pm6.8)\%$, with uncertainties accounting for the correlations between the two modes. Here, the first uncertainties are statistical, while the second are systematic. Additionally, we update the branching fraction of the decay $\psi(3686)\to\bar{\Xi}^{+}\Xi(1530)^{-}+\mathrm{c.c}$. The updated measurement is $\mathcal{B}(\psi(3686) \to \bar{\Xi}^{+}\Xi(1530)^{-}+\mathrm{c.c.})=(8.67\pm 0.52\pm0.58\pm0.57)\times 10^{-6}$, where the first uncertainty is statistical, the second is systematic related to event selection and the fit model, and the third is associated with the interference effect.
\end{abstract}

\maketitle

\section{Introduction}
The two-body hadronic decays of the $\psi(3686)$ resonance serve as a valuable test for perturbative quantum chromodynamics (QCD) and related calculations~\cite{refIntroQCD}. Decays of charmonium resonances to baryon-antibaryon pairs ($B\bar{B}$) are especially useful due to their simple topology. While many experimental results exist for decays $\psi(3686)\to B\bar{B}$~\cite{refpsipToXiXi, refpsipToXi(1530)0Xi(1530)0, refpsipToXiXi1530},
further investigation is still needed for decays to hyperon-antihyperon pairs such as $\psi(3686) \to \bar{\Xi}^{+}\Xi(1530)^{-}$.
According to SU(3)-flavor symmetry, decays of charmonium to $B\bar{B}$ pairs within the same multiplet are allowed,
while those involving different multiplets are suppressed~\cite{refSU3symmetry1, refSU3symmetry2}. However, SU(3) symmetry violation has been experimentally observed, including the decay $\psi(3686)\to\bar{\Xi}^{+}\Xi(1530)^{-}$. The BESIII Collaboration measured this decay in 2019 using $448.1\times10^{6}$ $\psi(3686)$ events~\cite{refpsipToXiXi1530}, employing single-baryon tagging to identify the decay and report the branching fractions (BFs) of the decays $\psi(3686)\to\bar{\Xi}^{+}\Xi(1530)^{-}$ and $\bar{\Xi}(1530)^{+}\Xi(1530)^{-}$.

As a well-established $\Xi^{*}$ resonance, the $\Xi(1530)$ state has been clearly observed in various production mechanisms including $K^{-}$-, $\Xi^{-}$-, and $\gamma$-induced reactions, as well as in the decays of charmed baryons~\cite{refIntro1Xi1530, refIntro2Xi1530, refIntro3Xi1530}. The properties are all fairly well known, and the spin-parity is determined to be $\frac{3}{2}^{+}$~\cite{refIntro2Xi1530, refJp1Xi1530, refJp2Xi1530}.
The $\Xi(1530)$ resonance was first observed in the $\Xi\pi$ decay mode, and its BF is listed as 100\% in the Particle Data Group (PDG)~\cite{refPDG}. However, there is no experimental measurement for the absolute BF of the decay $\Xi(1530)\to\Xi\pi$.

In this paper, based on $(2712.4\pm14.3)\times10^{6}$ $\psi(3686)$ events collected by the BESIII detector at BEPCII in 2009, 2012, and 2021~\cite{refpsipnumber}, we present the first measurement of the absolute BFs for the decays $\Xi(1530)^{-}\to\Xi^{0}\pi^{-}$ and $\Xi(1530)^{-}\to\Xi^{-}\pi^{0}$.
We also provide an updated BF measurement of $\psi(3686)\to\bar{\Xi}^{+}\Xi(1530)^{-}$ with improved precision.
Throughout the text, charge-conjugated modes are implied.

\section{BESIII detector and Monte Carlo simulations}
The BESIII detector~\cite{refBESIIIdetector} records symmetric $e^+e^-$ collisions
provided by the BEPCII collider~\cite{refBEPCII}
in the center-of-mass (CM) energy range from 1.84 to 4.95~GeV,
with a peak luminosity of $1.1 \times 10^{33}\;\text{cm}^{-2}\text{s}^{-1}$
achieved at $\sqrt{s} = 3.773~\text{GeV}$.
BESIII has collected large data samples in this energy region~\cite{refdatasample1, refdatasample2, refdatasample3}.
The cylindrical core of the BESIII detector covers 93\% of the full solid angle and consists of a helium-based
 multilayer drift chamber~(MDC), a plastic scintillator time-of-flight
system~(TOF), and a CsI(Tl) electromagnetic calorimeter~(EMC),
which are all enclosed in a superconducting solenoidal magnet
providing a 1.0~T magnetic field.
The solenoid is supported by an
octagonal flux-return yoke with resistive plate counter muon
identification modules (MUC) interleaved with steel.
The charged-particle momentum resolution at $1~{\rm GeV}/c$ is
$0.5\%$, and the specific energy loss (${\rm d}E/{\rm d}x$)
resolution is $6\%$ for electrons
from Bhabha scattering. The EMC measures photon energies with a
resolution of $2.5\%$ ($5\%$) at $1$~GeV in the barrel (end cap)
region. The time resolution in the TOF barrel region is 68~ps, while
that in the end cap region was 110~ps. The end cap TOF
system was upgraded in 2015 using a multigap resistive plate chamber
technology, providing a time resolution of
60~ps,
which benefits about 83\% of data used in this analysis~\cite{refetof1, refetof2, refetof3}.

Monte Carlo (MC) simulated data samples produced with a {\sc
geant4}-based~\cite{refgeant4} software package, which
includes the geometric description of the BESIII detector and the
detector response, are used to determine detection efficiencies
and to estimate backgrounds. The simulation models the beam
energy spread and initial-state radiation (ISR) in the $e^+e^-$
annihilations with the generator {\sc
kkmc}~\cite{refkkmc1, refkkmc2}.
The inclusive MC sample includes the production of the
$\psi(3686)$ resonance, the ISR production of the $J/\psi$, and
the continuum processes, which are generated with {\sc
kkmc}~\cite{refkkmc1, refkkmc2}.
All particle decays are modeled with {\sc
evtgen}~\cite{refevtgen1, refevtgen2} using BFs
either taken from the PDG~\cite{refPDG}, when available,
or otherwise estimated with {\sc lundcharm}~\cite{reflundcharm1, reflundcharm2}. Final-state radiation from charged final-state particles is included using the {\sc photos} package~\cite{refphotos}. The decay $\psi(3686)\to\bar{\Xi}^{+}\Xi(1530)^{-}$ is simulated with the J2BB3 model~\cite{refevtgen2}. The process $\bar{\Xi}^{+}\to\bar{\Lambda}\pi^{+}$ with $\bar{\Lambda}\to\bar{p}\pi^{+}$ is simulated with decay parameters fixed to the latest measurements~\cite{refdecaypar1}. The decay of $\Xi(1530)^{-}\to\Xi^{0(-)}\pi^{-(0)}$ is simulated with the phase space model, along with the decays of $\Xi^{0(-)}\to\Lambda\pi^{0(-)}$, $\Lambda\to p\pi^{-}$, and $\pi^{0}\to\gamma\gamma$ modeled with fixed decay parameters~\cite{refdecaypar1, refdecaypar2}. In addition, the mass and width of $\Xi(1530)^{-}$ in MC simulation are set according to our observations.

\section{Event selection and data analysis}
\subsection{Analysis Strategy}
The BF of $\psi(3686)\to\bar{\Xi}^{+}\Xi(1530)^{-}$ is measured using the single-tag (ST) technique. This involves reconstructing a $\bar{\Xi}^{+}$ baryon in its dominant decay mode $\bar{\Xi}^{+}\to\bar{\Lambda}\pi^{+}$ and identifying the $\Xi(1530)^{-}$ candidate on the recoil side. The absolute BF of $\Xi(1530)^{-}\to\Xi^{0(-)}\pi^{-(0)}$ is measured using the double-tag (DT) technique, which reconstructs the signal decay $\Xi(1530)^{-}\to\Xi^{0(-)}\pi^{-(0)}$ on top of the ST selection, in events from $\psi(3686)\to\bar{\Xi}^{+}\Xi(1530)^{-}$. When a $\bar{\Xi}^{+}$ hyperon is found, it is referred to as an ST candidate. If both the signal decay $\Xi(1530)^{-}\to\Xi^{0(-)}\pi^{-(0)}$ and the ST $\bar{\Xi}^{+}$ are detected, the event is designated as a DT event.
Therefore, for the exclusive signal decay $\Xi(1530)^{-}\to\Xi^{0}\pi^{-}$, the absolute BF is given by
\begin{equation}\label{eq:bf2}
\mathcal{B}(\Xi(1530)^{-} \to \Xi^{0}\pi^{-})=\frac{N_{\Xi^{0}\pi^{-}}\epsilon_{\mathrm{ST}}}{N_{\mathrm{ST}}\epsilon_{\Xi^{0}\pi^{-}}\mathcal{B}_{1}},
\end{equation}
where $N_{\Xi^{0}\pi^{-}}$ and $\epsilon_{\Xi^{0}\pi^{-}}$ are the DT signal yield and detection efficiency of this exclusive mode, $N_{\mathrm{ST}}$ and $\epsilon_{\mathrm{ST}}$ are the corresponding ST yield and efficiency, respectively, and $\mathcal{B}_{1}$ is the product of BFs of $\Xi^{0} \to \Lambda\pi^{0}$, $\Lambda \to p\pi^{-}$, and $\pi^{0} \to \gamma\gamma$ using values taken from the PDG~\cite{refPDG}. For the exclusive signal decay $\Xi(1530)^{-}\to\Xi^{-}\pi^{0}$, the absolute BF is given by
\begin{equation}\label{eq:bf1}
\mathcal{B}(\Xi(1530)^{-} \to \Xi^{-}\pi^{0})=\frac{N_{\Xi^{-}\pi^{0}}\epsilon_{\mathrm{ST}}}{N_{\mathrm{ST}}\epsilon_{\Xi^{-}\pi^{0}}\mathcal{B}_{2}},
\end{equation}
where $N_{\Xi^{-}\pi^{0}}$ and $\epsilon_{\Xi^{-}\pi^{0}}$ are the DT signal yield and detection efficiency of this exclusive mode, and $\mathcal{B}_{2}$ is the product of BFs of $\Xi^{-} \to \Lambda\pi^{-}$, $\Lambda \to p\pi^{-}$, and $\pi^{0} \to \gamma\gamma$~\cite{refPDG}.

\subsection{ST Analysis}
In the event selection, charged tracks detected in the MDC are required to be within $|\cos\theta|<0.93$, where the polar angle $\theta$ is defined with respect to the $z$-axis, which is the symmetry axis of the MDC. Due to the long lifetimes of hyperons $\Xi$ and $\Lambda$,
the charged tracks used in their reconstruction are not required to originate from the interaction point.
Particle identification~(PID) for charged tracks combines measurements of the specific ionization energy loss in the MDC and the flight time in the TOF to form likelihoods $\mathcal{L}(h)~(h=p,K,\pi)$ for each hadron $h$ hypothesis.
Tracks are identified as protons when the proton hypothesis has the greatest likelihood ($\mathcal{L}(p)>\mathcal{L}(K)$ and $\mathcal{L}(p)>\mathcal{L}(\pi)$), while charged pions are identified requiring $\mathcal{L}(\pi)>\mathcal{L}(p)$ and $\mathcal{L}(\pi)>\mathcal{L}(K)$. Events with at least two positively charged pions and one antiproton are selected as ST $\bar{\Xi}^{+}$ candidates.

The decay chain $\bar{\Xi}^{+}\to\bar{\Lambda}\pi^{+}\to\bar{p}\pi^{+}\pi^{+}$ is reconstructed by performing a primary vertex fit and a secondary vertex fit, taking into account the flight paths of the hyperon $\bar{\Xi}^{+}$. If there are multiple combinations, the combination with the minimum value of $\delta\equiv|M_{\bar{p}\pi^{+}}-m_{\bar{\Lambda}}|+|M_{\bar{p}\pi^{+}\pi^{+}}-m_{\bar{\Xi}^{+}}|$ is retained, where $M_{\bar{p}\pi^{+}\pi^{+}} ~ (M_{\bar{p}\pi^{+}})$ denotes the invariant mass of $\bar{p}\pi^{+}\pi^{+} ~ (\bar{p}\pi^{+})$ obtained from the vertex fit, and $m_{\bar{\Xi}^{+}} ~ (m_{\bar{\Lambda}})$ refers to the nominal mass of $\bar{\Xi}^{+}(\bar{\Lambda})$~\cite{refPDG}. The candidate events are further required to satisfy $|M_{\bar{p}\pi^{+}}-m_{\bar{\Lambda}}|\leq 5\ \mathrm{MeV}/c^{2}$ and $|M_{\bar{p}\pi^{+}\pi^{+}}-m_{\bar{\Xi}^{+}}|\leq 9\ \mathrm{MeV}/c^{2}$, which correspond to about three times the respective mass resolution. The decay length of $\bar{\Xi}^{+}$ obtained from the vertex fit is required to be positive to further suppress background events.

To suppress background events from the decay $\psi(3686)\to\pi^{+}\pi^{-}J/\psi$, we require $|M_{\pi^{+}\pi^{-}}^{\mathrm{recoil}}-m_{J/\psi}|\geq 7.5\ \mathrm{MeV}/c^{2}$, where $M_{\pi^{+}\pi^{-}}^{\mathrm{recoil}}$ is the recoil mass of the $\pi^{+}\pi^{-}$ system, and $m_{J/\psi}$ is the nominal mass of the $J/\psi$ meson~\cite{refPDG}. The requirement corresponds to about two and a half times the mass resolution.

The ST signal is identified with the recoil mass of $\bar{\Xi}^{+}$, defined as
\begin{equation}\label{eq:recXi}
M_{\bar{\Xi}^{+}}^{\mathrm{recoil}}\equiv\sqrt{(E_{\mathrm{CM}}-E_{\bar{\Lambda}\pi^{+}})^{2}-|\vec{p}_{\bar{\Lambda}\pi^{+}}|^{2}},
\end{equation}
where $E_{\mathrm{CM}}$ is the CM energy of the $e^{+}e^{-}$ system, and $E_{\bar{\Lambda}\pi^{+}}$ and $\vec{p}_{\bar{\Lambda}\pi^{+}}$ are the energy and momentum of the selected $\bar{\Xi}^{+}$ candidate in the CM system, respectively. The recoil mass of $\bar{\Xi}^{+}$ is required to be within the $\Xi(1530)^{-}$ signal range of $[1.43,1.65]\ \mathrm{GeV}/c^{2}$ to determine the signal yield.

\begin{figure}
\begin{overpic}[scale=.34]{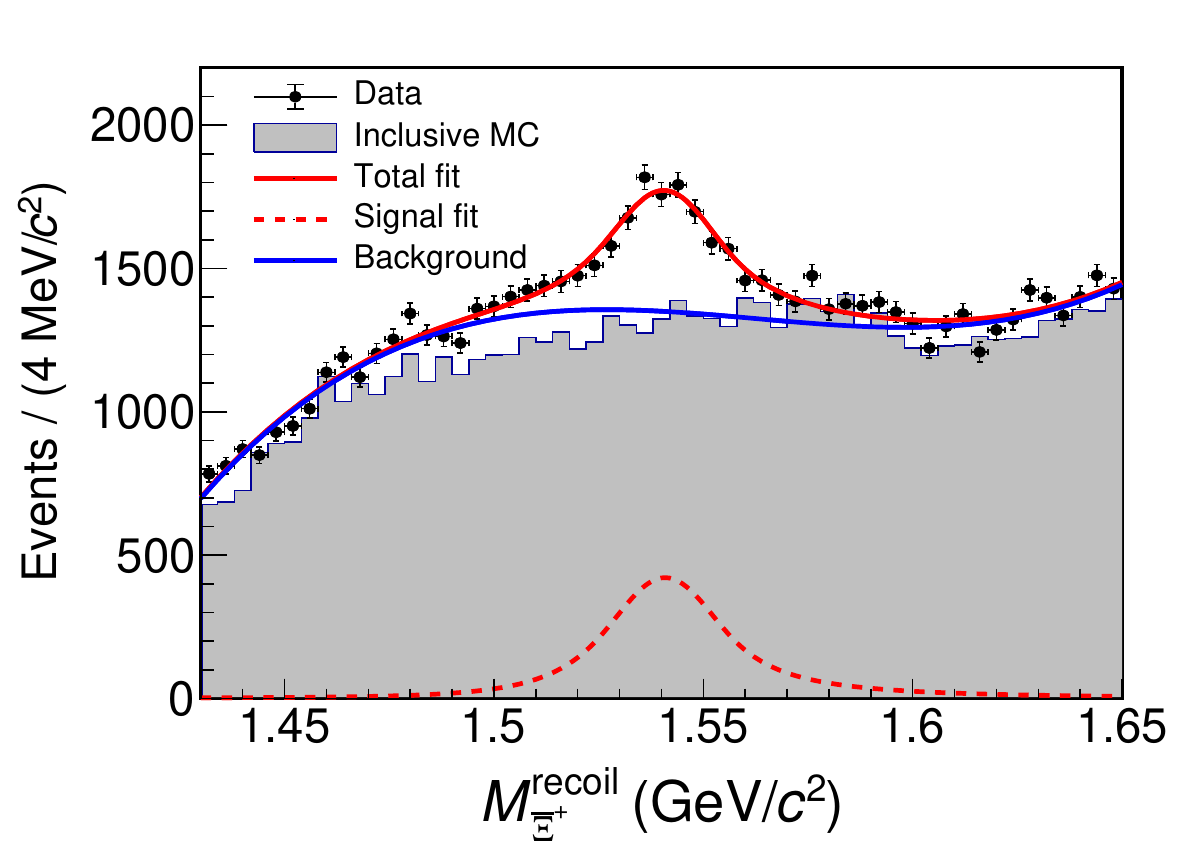}
\end{overpic}
\caption{\label{fig:ST_recXibar}Fit to the $M_{\bar{\Xi}^{+}}^{\mathrm{recoil}}$ distribution of the ST $\bar{\Xi}^{+}$ candidates. Data are shown as dots with error bars. The solid red and blue curves are the total fit and the background contribution, respectively. The dashed red curve is the fitted $\Xi(1530)^{-}$ signal. The shaded histogram represents the background shape of the inclusive MC sample.}
\end{figure}

After applying all the aforementioned ST selection criteria, the remaining background events are investigated using the inclusive MC sample with the {\sc topoana}~\cite{reftopo} package. In the signal region of the $\Xi(1530)^{-}$ mass, no peaking background events are observed, as shown in Fig.~\ref{fig:ST_recXibar}.

We also investigate the background events from the continuum production with an off-resonance data sample taken at the CM energy of 3.650 GeV that corresponds to an integrated luminosity of $(445.49\pm4.02)\ \mathrm{pb}^{-1}$~\cite{refcontinuumluminosity, refpsipnumber}. The surviving events do not contribute to the peak position of $\Xi(1530)^{-}$ in the $M_{\bar{\Xi}^{+}}^{\mathrm{recoil}}$ spectrum.

Considering the luminosity and cross section factors,  we estimate the number of background events from the continuum process by
\begin{equation}\label{eq:continuum}
N_{\mathrm{con.}}=\sigma_{c}\cdot\mathcal{L}_{3.686}\cdot\epsilon_{\mathrm{con.}}\cdot\mathcal{B}(\bar{\Xi}^{+}\to\bar{\Lambda}\pi^{+})\cdot\mathcal{B}(\bar{\Lambda}\to\bar{p}\pi^{+}),
\end{equation}
where $\sigma_{c}=(81.96\pm6.09)\ \mathrm{fb}$ is the cross section of $e^{+}e^{-}\to\bar{\Xi}^{+}\Xi(1530)^{-}$ at the CM energy of $3.686\ \mathrm{GeV}$, obtained from the fitted curve of measured cross sections at different CM energy points~\cite{refcontinuumcrosssection}; $\mathcal{L}_{3.686}=(3877.05\pm32.10)\ \mathrm{pb}^{-1}$ is the integrated luminosity at 3.686 $\mathrm{GeV}$~\cite{refpsipnumber}; $\epsilon_{\mathrm{con.}}$ is the detection efficiency of the continuum process; $\mathcal{B}(\bar{\Xi}^{+} \to \bar{\Lambda}\pi^{+})$ and $\mathcal{B}(\bar{\Lambda} \to \bar{p}\pi^{+})$ are the BFs of $\bar{\Xi}^{+} \to \bar{\Lambda}\pi^{+}$ and $\bar{\Lambda} \to \bar{p}\pi^{+}$~\cite{refPDG}, respectively. Therefore, the number of background events from the continuum process is estimated to be $N_{\mathrm{con.}}=59\pm4$.

The $M_{\bar{\Xi}^{+}}^{\mathrm{recoil}}$ distribution of surviving events is shown in Fig.~\ref{fig:ST_recXibar}. The $\Xi(1530)^{-}$ signal yield based on the ST $\bar{\Xi}^{+}$ sample is determined using an unbinned extended maximum likelihood fit to the $M_{\bar{\Xi}^{+}}^{\mathrm{recoil}}$ distribution, where the $\Xi(1530)^{-}$ signal is modeled by a Breit-Wigner function convolved with a double-sided Crystal Ball function~\cite{refCB} to account for the detector resolution, and the background is modeled by a third-order Chebyshev polynomial function. In the fit, the parameters of the Breit-Wigner function are left free, while the parameters of the double-sided Crystal Ball function are fixed to the values determined from signal MC simulation. The fit result is shown in Fig.~\ref{fig:ST_recXibar}.

After subtracting the contribution from the continuum process, the total ST $\Xi(1530)^{-}$ signal yield is $N_{\mathrm{ST}}=4392\pm261$, and the ST detection efficiency from MC simulation is $\epsilon_{\mathrm{ST}}=(29.17\pm0.06)\%$.

The BF of $\psi(3686)\to\bar{\Xi}^{+}\Xi(1530)^{-}$ is calculated as
\begin{equation}\begin{split}\label{eq:cal_STbf}
&\mathcal{B}(\psi(3686) \to \bar{\Xi}^{+}\Xi(1530)^{-})\\
&=\frac{N_{\mathrm{ST}}}{\epsilon_{\mathrm{ST}} N_{\psi(3686)}\mathcal{B}(\bar{\Xi}^{+} \to \bar{\Lambda}\pi^{+})\mathcal{B}(\bar{\Lambda} \to \bar{p}\pi^{+})} ,
\end{split}\end{equation}
where $N_{\psi(3686)}=(2712.4\pm14.3)\times10^{6}$ is the total number of $\psi(3686)$ events~\cite{refpsipnumber}.

\subsection{DT Analysis}
Based on the surviving ST candidates, the signals $\Xi(1530)^{-}\to\Xi^{0}\pi^{-}$ and $\Xi(1530)^{-}\to\Xi^{-}\pi^{0}$ are reconstructed with the remaining tracks recoiling against the ST $\bar{\Xi}^{+}$ candidates. Photon candidates are identified from showers in the EMC. The deposited energy of each shower is required to be greater 25 MeV in the barrel region ($|\cos\theta\leq 0.80|$) and greater than 50 MeV in the end cap region ($0.86\leq|\cos\theta|\leq 0.92$). To suppress electronic noise and energy deposits unrelated to the event, the difference between the EMC time and the event start time is required to be within $[0,700]\ \mathrm{ns}$. The number of photon candidates is required to be at least two. The $\pi^{0}$ candidates are reconstructed by considering all photon-pair combinations, and the photon pairs falling within the $\pi^{0}$ mass window $[0.11,0.16]\ \mathrm{GeV}/c^{2}$ are retained as $\pi^{0}$ candidates. We further require a candidate event to contain at least two negatively charged tracks identified as pions and one positively charged track identified as a proton.

Similar to the ST reconstruction, each $\Lambda$ candidate is reconstructed from a $p\pi^{-}$ pair by a vertex fit, requiring the invariant mass to be within $5\ \mathrm{MeV}/c^{2}$ of the nominal $\Lambda$ mass. Additionally, we compile a list of the $\bar{p}\pi^{+}\pi^{+}$ combinations for $\bar{\Xi}^{+}$ that satisfy the vertex fit criteria, mass window, and decay length requirements during ST reconstruction, without selecting the best combination based on $\delta=|M_{\bar{p}\pi^{+}}-m_{\bar{\Lambda}}|+|M_{\bar{p}\pi^{+}\pi^{+}}-m_{\bar{\Xi}^{+}}|$. To suppress background, a four-constraint (4C) kinematic fit is performed to the decay product of $\bar{\Xi}^{+}$ and $\Lambda$, $\pi^{+}\pi^{+}\pi^{-}p\bar{p}$, together with an additional $\pi^{-}$ and a $\pi^{0}$ from the $\pi^{0}$ candidate list. This fit enforces energy-momentum conservation, constraining the CM energy of the $e^{+}e^{-}$ system. If multiple combinations are present, the $\pi^{+}\pi^{+}\pi^{-}\pi^{-}\pi^{0}p\bar{p}$ combination with the minimum $\chi_{\mathrm{4C}}^{2}$ is selected as the best candidate, where $\chi_{\mathrm{4C}}^{2}$ is the goodness of the 4C kinematic fit for the $\pi^{+}\pi^{+}\pi^{-}\pi^{-}\pi^{0}p\bar{p}$ combination.
To suppress backgrounds containing only one photon, in contrast to the two-photon topology of the signal events,
an analogous 4C kinematic fit is also applied to the $\pi^{+}\pi^{+}\pi^{-}\pi^{-}p\bar{p}\gamma$ combinations, yielding a minimum $\chi_{\mathrm{4C,1\gamma}}^{2}$ value.
We require $\chi_{\mathrm{4C}}^{2}<\chi_{\mathrm{4C},1\gamma}^{2}$ to improve the signal purity.

The $\Xi$ candidates are reconstructed by considering all $\Lambda\pi$ combinations in the final state. For the $\Xi^{0}$ candidates, the $\Lambda\pi^{0}$ combination with the minimum value of $|M_{p\pi^{-}\pi^{0}}-m_{\Xi^{0}}|$ is retained, where $M_{p\pi^{-}\pi^{0}}$ denotes the invariant mass of $p\pi^{-}\pi^{0}$ obtained from the 4C kinematic fit, and $m_{\Xi^{0}}$ is the nominal mass of $\Xi^{0}$~\cite{refPDG}. For the $\Xi^{-}$ and $\Xi(1530)^{-}$ candidates, which decay into $p\pi^{-}\pi^{-}$ and $p\pi^{-}\pi^{-}\pi^{0}$, respectively, only one combination for each of them is available in the final state.

To veto background events from the decay $\psi(3686)\to\eta J/\psi$, $\eta\to\pi^{+}\pi^{-}\pi^{0}$, we require the recoil mass of the $\pi^{+}\pi^{-}\pi^{0}$ system, $M_{\pi^{+}\pi^{-}\pi^{0}}^{\mathrm{recoil}}$,
to satisfy
$|M_{\pi^{+}\pi^{-}\pi^{0}}^{\mathrm{recoil}}-m_{J/\psi}|\geq15\ \mathrm{MeV}/c^{2}$. To veto the $\psi(3686)\to\pi^{0}h_{c}$ background events, we impose the condition $|M_{\pi^{0}}^{\mathrm{recoil}}-m_{h_{c}}|\geq10\ \mathrm{MeV}/c^{2}$, where $M_{\pi^{0}}^{\mathrm{recoil}}$ is the recoil mass of $\pi^{0}$, and $m_{h_{c}}$ is the nominal mass of $h_{c}$~\cite{refPDG}. These requirements correspond to about two and a half times the mass resolution.

After applying all DT selection criteria, the remaining background candidates are studied using the $\psi(3686)$ inclusive MC sample with the {\sc topoana} package~\cite{reftopo}. Most of the background events share the same final states as the signal events and are difficult to suppress. No significant peaking background is observed in the $\Xi(1530)^{-}$ invariant mass spectrum.

\begin{figure*}
\begin{overpic}[scale=.34]{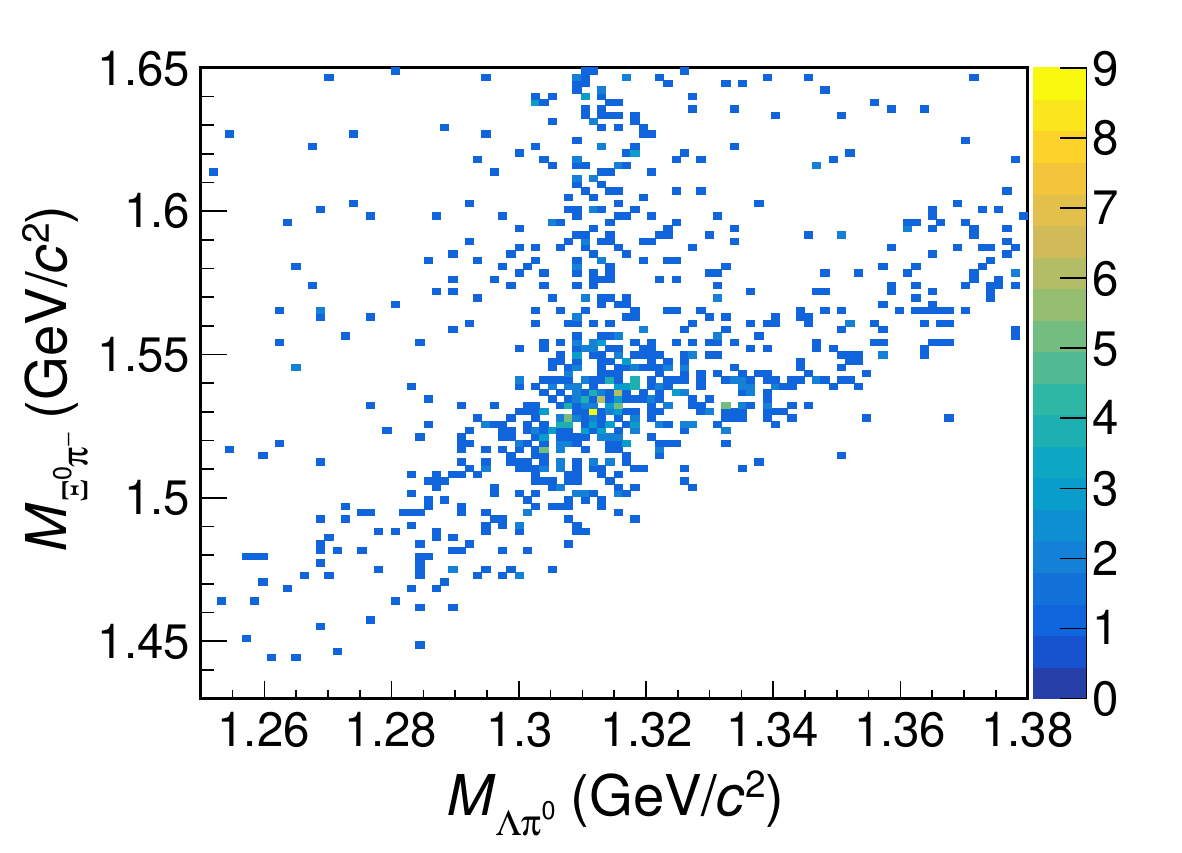}
\end{overpic}
\begin{overpic}[scale=.34]{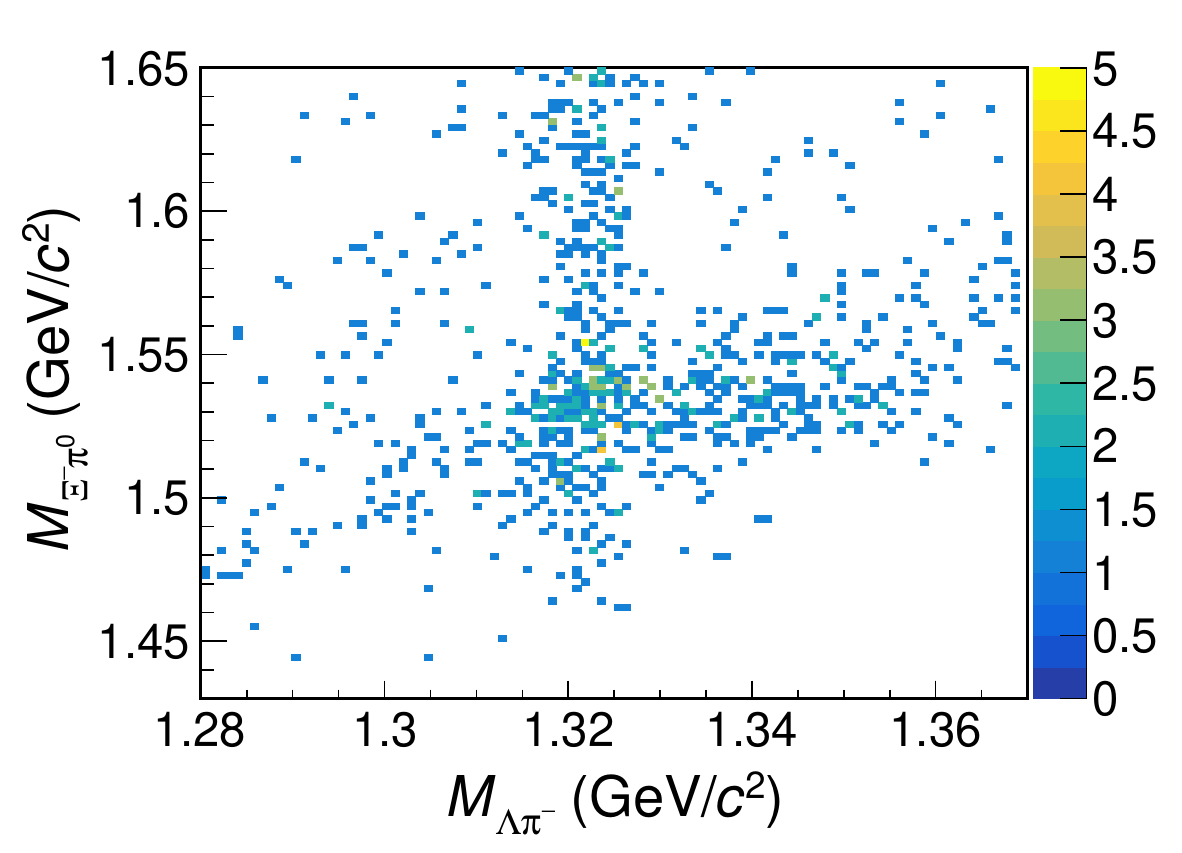}
\end{overpic}
\caption{The 2-D distributions of $M_{\Xi^{0}}$ versus $M_{\Xi(1530)^{-}}$ (left) and $M_{\Xi^{-}}$ versus $M_{\Xi(1530)^{-}}$ (right) from data.}
\label{fig:DT_2d_scatterplot}
\end{figure*}

For the DT mode $\Xi(1530)^{-}\to\Xi^{0}\pi^{-}$, the remaining events can be classified into four categories based on their distinct peaking characteristics:
(1) a $\Xi^{0}$ signal with a $\Xi(1530)^{-}$ signal from signal events; (2) a $\Xi^{0}$ signal with non-$\Xi(1530)^{-}$; (3) non-$\Xi^{0}$ with non-$\Xi(1530)^{-}$; and (4) non-$\Xi^{0}$ with a $\Xi(1530)^{-}$ signal from the isospin decay mode $\Xi(1530)^{-}\to\Xi^{-}\pi^{0}$,
which shares the same final state as the signal events but differs in the intermediate resonance state. While the fourth category constitutes background for the DT mode $\Xi(1530)^{-}\to\Xi^{0}\pi^{-}$, it serves as a signal in the DT mode $\Xi(1530)^{-}\to\Xi^{-}\pi^{0}$. The two-dimensional (2-D) distribution of $M_{\Xi^{0}}$ versus $M_{\Xi(1530)^{-}}$ from data is presented in Fig.~\ref{fig:DT_2d_scatterplot}, where $M_{\Xi^{0}}$ and $M_{\Xi(1530)^{-}}$ are the invariant masses of $\Xi^{0}$ and $\Xi(1530)^{-}$, respectively.

For the DT mode $\Xi(1530)^{-}\to\Xi^{-}\pi^{0}$, the remaining events are categorized into four types using the same method as above. The 2-D distribution of $M_{\Xi^{-}}$ versus $M_{\Xi(1530)^{-}}$ from data is presented in Fig.~\ref{fig:DT_2d_scatterplot}, where $M_{\Xi^{-}}$ is the invariant mass of $\Xi^{-}$.

To extract the DT signal yield, a simultaneous unbinned extended maximum likelihood fit is performed to the 2-D distribution of $M_{\Xi^{0}}$ versus $M_{\Xi(1530)^{-}}$ from $\Xi(1530)^{-}\to\Xi^{0}\pi^{-}$ and the 2-D distribution of $M_{\Xi^{-}}$ versus $M_{\Xi(1530)^{-}}$ from $\Xi(1530)^{-}\to\Xi^{-}\pi^{0}$,
incorporating the isospin correlation
between decays $\Xi(1530)^{-}\to\Xi^{0}\pi^{-}$ and $\Xi(1530)^{-}\to\Xi^{-}\pi^{0}$.

In the simultaneous fit, the 2-D fit function for the DT mode $\Xi(1530)^{-}\to\Xi^{0}\pi^{-}$ is constructed as
\begin{equation}\begin{split}\label{eq:2dfit_function_dt2}
f(\Xi^{0},\Xi(1530)^{-})=N_{\Xi^{0}\pi^{-}}s_{1}(\Xi^{0})s_{2}(\Xi(1530)^{-})\\
+N_{\mathrm{bkg1}}s_{1}(\Xi^{0})b_{2}(\Xi(1530)^{-})\\
+N_{\mathrm{bkg2}}b_{1}(\Xi^{0})b_{2}(\Xi(1530)^{-})\\
+N_{\mathrm{bkg3}}bs_{\mathrm{2\text{-}D}}(\Xi^{0},\Xi(1530)^{-}).
\end{split}\end{equation}
Here, $N_j$ denotes the corresponding yield. The functions $s_{1}(\Xi^{0})$ and $s_{2}(\Xi(1530)^{-})$ denote the signal probability density functions (PDFs) for the $M_{\Xi^{0}}$ and $M_{\Xi(1530)^{-}}$ distributions, respectively, modeled by the MC signal shape convolved with a Gaussian function. The corresponding background PDFs, $b_{1}(\Xi^{0})$ and $b_{2}(\Xi(1530)^{-})$, are modeled by a second-order Chebyshev polynomial function and an anti-ARGUS function~\cite{refargus}, respectively, obtained from the inclusive MC sample. The last component, $bs_{\mathrm{2\text{-}D}}(\Xi^{0},\Xi(1530)^{-})$ represents a 2-D background PDF for the 2-D distribution of $M_{\Xi^{0}}$ versus $M_{\Xi(1530)^{-}}$,
extracted directly from the exclusive MC sample of $\Xi(1530)^{-}\to\Xi^{-}\pi^{0}$.
Based on the isospin relation and phase space factors, the ratio $r\equiv\frac{\mathcal{B}(\Xi(1530)^{-}\to\Xi^{0}\pi^{-})}{\mathcal{B}(\Xi(1530)^{-}\to\Xi^{-}\pi^{0})}$ is set to 2.06. Combining the relationship among yields of signal events, detection efficiencies, and BFs, $N_{\mathrm{bkg3}}$ is related to $N_{\Xi^{0}\pi^{-}}$ in the fit shown as
\begin{equation}\begin{split}\label{eq:simufit_related1}
N_{\mathrm{bkg3}}&=\frac{1}{r}\frac{\epsilon_{\mathrm{bkg3}}\mathcal{B}(\Xi^{-}\to\Lambda\pi^{-})}{\epsilon_{\Xi^{0}\pi^{-}}\mathcal{B}(\Xi^{0}\to\Lambda\pi^{0})}N_{\Xi^{0}\pi^{-}}=0.49N_{\Xi^{0}\pi^{-}},
\end{split}\end{equation}
where $\epsilon_{\Xi^{0}\pi^{-}}=(5.51\pm0.03)\%$ is the DT detection efficiency determined by MC simulation, $\epsilon_{\mathrm{bkg3}}=(5.54\pm0.03)\%$ is the contamination rate obtained from the exclusive MC sample of $\Xi(1530)^{-}\to\Xi^{-}\pi^{0}$ after applying the selection criteria of the DT mode $\Xi(1530)^{-}\to\Xi^{0}\pi^{-}$, and $\mathcal{B}(\Xi^{0}\to\Lambda\pi^{0})$ and $\mathcal{B}(\Xi^{-}\to\Lambda\pi^{-})$ denote the BFs of $\Xi^{0}\to\Lambda\pi^{0}$ and $\Xi^{-}\to\Lambda\pi^{-}$~\cite{refPDG}, respectively.

The 2-D fit function for the DT mode $\Xi(1530)^{-}\to\Xi^{-}\pi^{0}$ is constructed analogous to that of the decay mode $\Xi(1530)^{-}\to\Xi^{0}\pi^{-}$. The number of signal events is represented by $N_{\Xi^{-}\pi^{0}}$, and the number of background events that peak in the $M_{\Xi(1530)^{-}}$ distribution but not in the $M_{\Xi^{-}}$ distribution is denoted by $N_{\mathrm{bkg3}}^{\prime}$. In the fit, $N_{\Xi^{-}\pi^{0}}$ and $N_{\mathrm{bkg3}}^{\prime}$ are related to $N_{\Xi^{0}\pi^{-}}$ using
\begin{equation}\begin{split}\label{eq:simufit_related2}
N_{\Xi^{-}\pi^{0}}&=\frac{1}{r}\frac{\epsilon_{\Xi^{-}\pi^{0}}\mathcal{B}(\Xi^{-}\to\Lambda\pi^{-})}{\epsilon_{\Xi^{0}\pi^{-}}\mathcal{B}(\Xi^{0}\to\Lambda\pi^{0})}N_{\Xi^{0}\pi^{-}}=0.51N_{\Xi^{0}\pi^{-}},\\
N_{\mathrm{bkg3}}^{\prime}&=\frac{\epsilon_{\mathrm{bkg3}}^{\prime}}{\epsilon_{\Xi^{0}\pi^{-}}}N_{\Xi^{0}\pi^{-}}=0.93N_{\Xi^{0}\pi^{-}},
\end{split}\end{equation}
where $\epsilon_{\Xi^{-}\pi^{0}}=(5.72\pm0.03)\%$ is the DT detection efficiency, and $\epsilon_{\mathrm{bkg3}}^{\prime}=(5.13\pm0.03)\%$ is the contamination rate obtained from the exclusive MC sample of $\Xi(1530)^{-}\to\Xi^{0}\pi^{-}$ after applying the selection criteria of the DT mode $\Xi(1530)^{-}\to\Xi^{-}\pi^{0}$.

\begin{figure*}
\begin{overpic}[scale=.34]{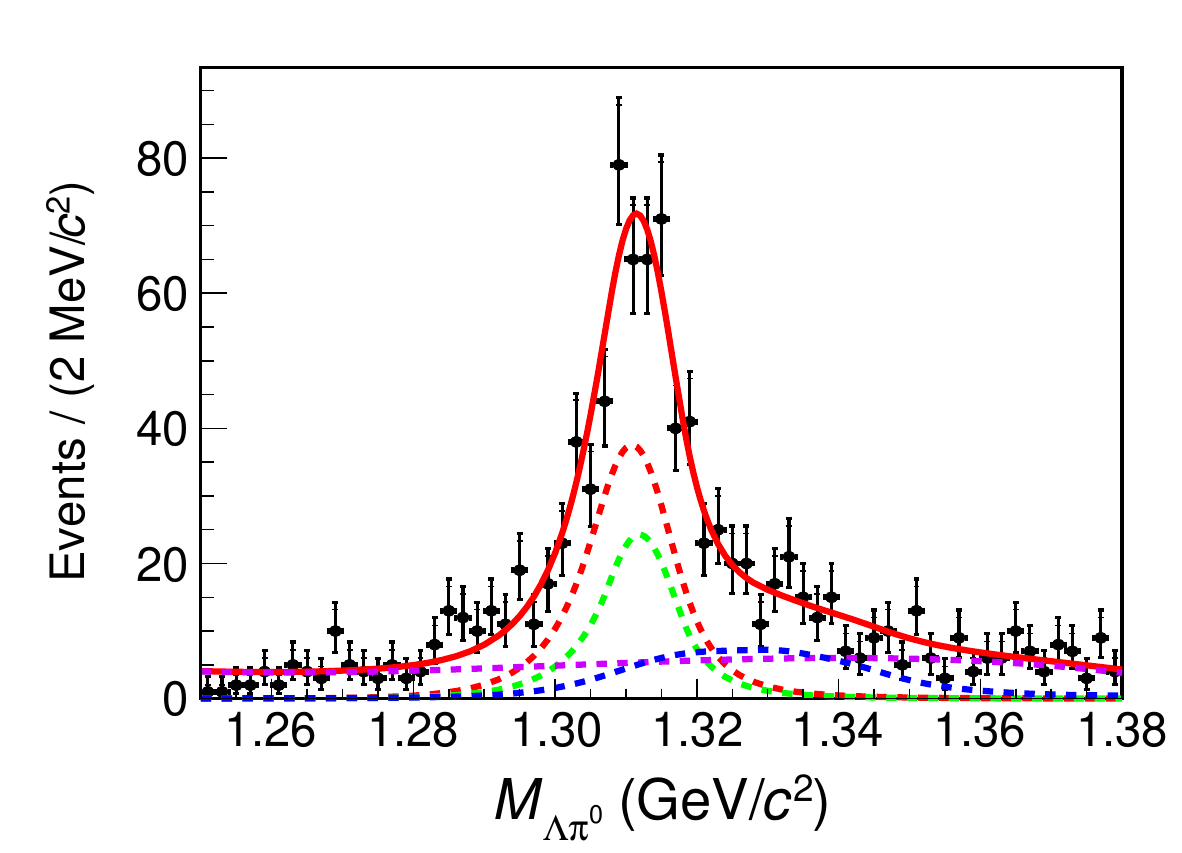}
\end{overpic}
\begin{overpic}[scale=.34]{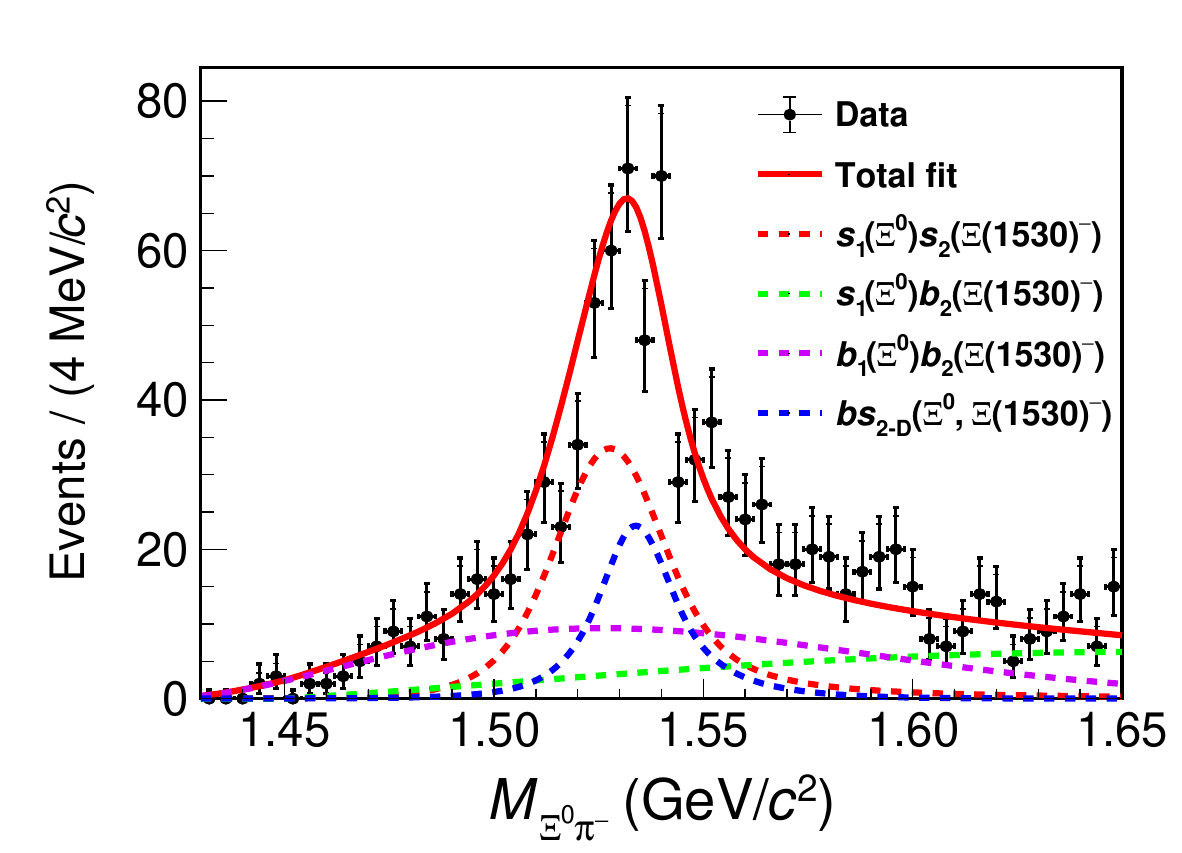}
\end{overpic}
\begin{overpic}[scale=.34]{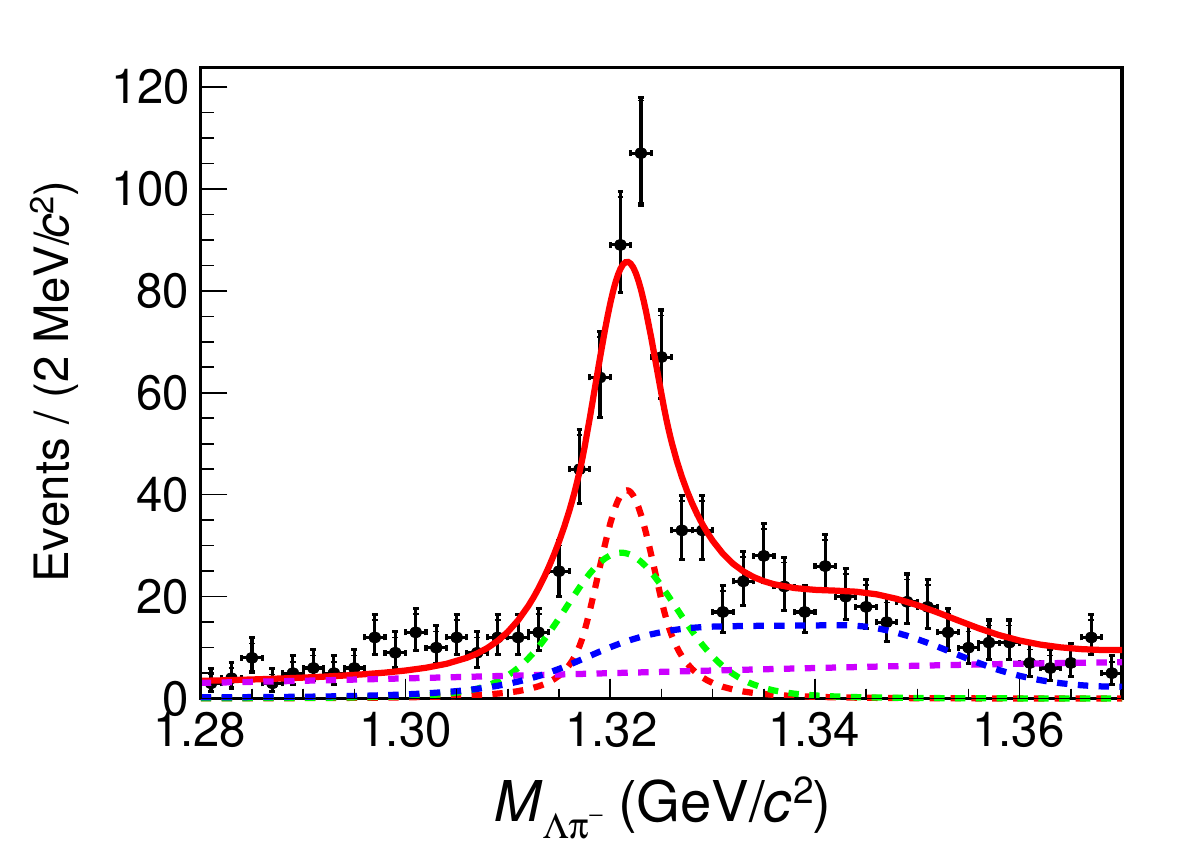}
\end{overpic}
\begin{overpic}[scale=.34]{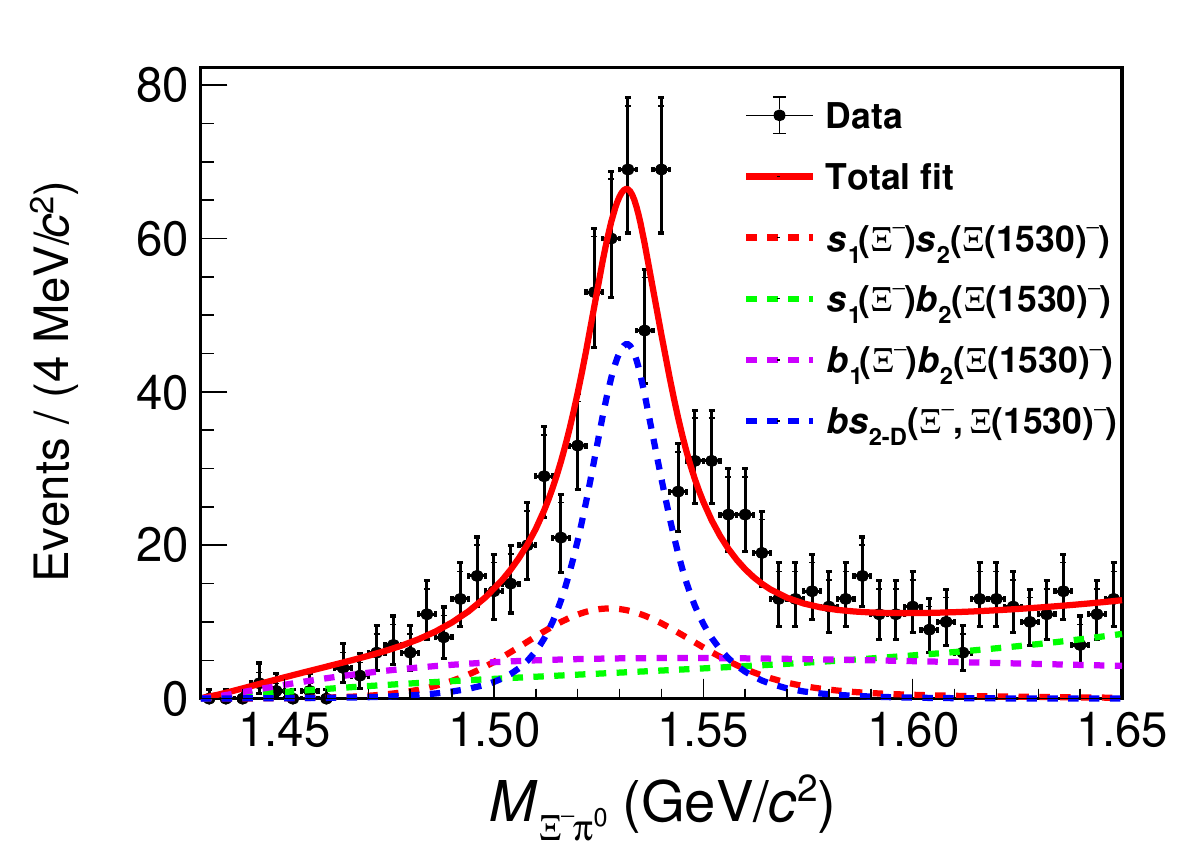}
\end{overpic}
\caption{Projections of simultaneous fit result for data. The top row shows the projections for $\Xi(1530)^{-}\to\Xi^{0}\pi^{-}$ and the bottom row shows those for $\Xi(1530)^{-}\to\Xi^{-}\pi^{0}$. Dots with error bars are for data. The red solid curve is for the total fit, the red dashed curve is for the signal shape, and the other dashed curves are for the three kinds of background.}
\label{fig:DT_simufit}
\end{figure*}

Therefore, in the simultaneous fit, $N_{\mathrm{bkg3}}$, $N_{\Xi^{-}\pi^{0}}$, and $N_{\mathrm{bkg3}}^{\prime}$ are related to $N_{\Xi^{0}\pi^{-}}$ according to Eq.~\eqref{eq:simufit_related1} and Eq.~\eqref{eq:simufit_related2}. Rather than treating the two signal yields as independent parameters, we take the total signal yield, $N_{\mathrm{tot}} \equiv N_{\Xi^{0}\pi^{-}}+N_{\Xi^{-}\pi^{0}}$, as the free parameter. The individual yields are fixed to $N_{\Xi^{0}\pi^{-}}=0.66N_{\mathrm{tot}}$ and $N_{\Xi^{-}\pi^{0}}=0.34N_{\mathrm{tot}}$ based on Eq.~\eqref{eq:simufit_related2}. The simultaneous fit, shown in Fig.~\ref{fig:DT_simufit}, yields $N_{\mathrm{tot}}=483\pm21$. The signal yields of the DT modes $\Xi(1530)^{-}\to\Xi^{0}\pi^{-}$ and $\Xi(1530)^{-}\to\Xi^{-}\pi^{0}$ are therefore $N_{\Xi^{0}\pi^{-}}=321\pm14$ and $N_{\Xi^{-}\pi^{0}}=162\pm7$,
respectively, where the statistical uncertainties are propagated from the uncertainty of $N_{\mathrm{tot}}$. 

The combined BF of $\Xi(1530)^{-} \to (\Xi\pi)^{-}$ is calculated using Eq.~\eqref{eq:bftot}, shown as
\begin{equation}\label{eq:bftot}
\mathcal{B}(\Xi(1530)^{-} \to (\Xi\pi)^{-})=\frac{N_{\mathrm{tot}}\epsilon_{\mathrm{ST}}}{N_{\mathrm{ST}}(f_1\epsilon_{\Xi^{0}\pi^{-}}\mathcal{B}_{1}+f_2\epsilon_{\Xi^{-}\pi^{0}}\mathcal{B}_{2})}.
\end{equation}
Here, the factors $f_1\equiv \frac{\mathcal{B}(\Xi(1530)^{-} \to \Xi^{0}\pi^{-})}{\mathcal{B}(\Xi(1530)^{-} \to (\Xi\pi)^{-})}$ and $f_2\equiv \frac{\mathcal{B}(\Xi(1530)^{-} \to \Xi^{-}\pi^{0})}{\mathcal{B}(\Xi(1530)^{-} \to (\Xi\pi)^{-})}$ are determined to be 0.67 and 0.33, respectively, based on the ratio $r=2.06$. The absolute BFs of $\Xi(1530)^{-} \to \Xi^{0}\pi^{-}$ and $\Xi(1530)^{-}\to\Xi^{-}\pi^{0}$ are calculated using Eq.~\eqref{eq:bf2} and Eq.~\eqref{eq:bf1}, respectively. The signal yields in data, along with the BFs ($\mathcal{B}$) for the decay modes analyzed, are summarized in Table~\ref{table:bfresult_summary}.

\begin{table*}[htbp]
\centering
\caption{Summary of the signal yields in data and measured BFs. The first uncertainties are statistical, and the second are systematic. The symbol $\mathcal{B}$ denotes the BF. For the BF of $\psi(3686)\to\bar{\Xi}^{+}\Xi(1530)^{-}$, the first uncertainty is statistical, the second is systematic related to event selection and the fit model, and the third is systematic related to the interference effect.}
\label{table:bfresult_summary}

\begin{tabular}{
    r @{${}\to{}$} l      @{\hspace{4ex}}
    r @{${}\pm{}$} l      @{\hspace{4ex}}
    r @{${}\pm{}$} l @{${}\pm{}$} l
}
\hline\hline
\multicolumn{2}{c}{Mode} & \multicolumn{2}{c}{Signal yield} & \multicolumn{3}{c}{$\mathcal{B}$} \\ \hline
$\psi(3686)$ & $\bar{\Xi}^{+}\Xi(1530)^{-}$ &
4392 & 261 &
$(8.67$ & $0.52$ & $0.58\pm0.57)\times10^{-6}$ \\

$\Xi(1530)^{-}$ & $\Xi^{0}\pi^{-}$ &
321 & 14 &
$(61.4$ & $4.5$ & $4.6)\%$ \\

$\Xi(1530)^{-}$ & $\Xi^{-}\pi^{0}$ &
162 & 7 &
$(29.7$ & $2.2$ & $2.2)\%$ \\

$\Xi(1530)^{-}$ & $(\Xi\pi)^{-}$ &
483 & 21 &
$(91.1$ & $6.7$ & $6.8)\%$ \\
\hline\hline
\end{tabular}
\end{table*}

\section{Systematic uncertainties}

In the BF measurements, systematic uncertainties arise from the following aspects: event selection, the fit model, and the interference effect between the resonance decay and continuum process. For the two DT modes, uncertainties related to the $\bar{\Xi}^{+}$ reconstruction mostly cancel, while those associated with the $\Xi(1530)^{-}$ reconstruction are taken into account. The sources of systematic uncertainties are summarized in Table~\ref{table:sysuncertainty_summary}, where the total systematic uncertainty is the quadratic sum of the individual components.

The uncertainty related to the $\bar{\Xi}^{+}(\Xi^{-})$ reconstruction is estimated with the control sample of $\psi(3686)\to\Xi^{-}\bar{\Xi}^{+}$. The efficiency difference between data and MC simulation, 1.3\%, is assigned as the systematic uncertainty.

The uncertainty related to the $\Lambda(\bar{\Lambda})$ reconstruction is also estimated with the control sample of $\psi(3686)\to\Xi^{-}\bar{\Xi}^{+}$. The efficiency difference between data and MC simulation, 1.0\%, is assigned as the systematic uncertainty.

The uncertainty associated with the $\pi^{-}$ tracking and the PID is estimated to be 1.4\% by studying a control sample of $J/\psi\to\pi^{+}\pi^{-}p\bar{p}$~\cite{refpsipToXiXi1530}.

The uncertainty related to the $\pi^{0}$ reconstruction arises from two sources: the photon reconstruction and the $\pi^{0}$ mass window. The uncertainty due to the photon reconstruction is assigned to be 1\% per photon, by studying control samples of $J/\psi\to\rho^{0}\pi^{0}$ and $e^{+}e^{-}\to\gamma\gamma$~\cite{refphoton_rec_eff}. The uncertainty due to the $\pi^{0}$ mass window is estimated to be 0.05\% by comparing the efficiencies between data and MC simulation.

The uncertainties related to the mass windows requirements for $\bar{\Lambda}$, $\bar{\Xi}^{+}$, and $\Lambda$ are estimated from the data-MC efficiency difference, to be 0.04\%, 0.03\%, and 0.1\%, respectively.

The uncertainties associated with the 4C kinematic fit are determined from the efficiency differences before and after the tracking helix correction~\cite{refkmfit_coreff}, which are 0.04\% and 0.1\% for the DT modes $\Xi(1530)^{-}\to\Xi^{0}\pi^{-}$ and $\Xi(1530)^{-}\to\Xi^{-}\pi^{0}$, respectively.

The uncertainties associated with the requirement of $\chi_{\mathrm{4C}}^{2}<\chi_{\mathrm{4C},1\gamma}^{2}$ are estimated using the Barlow test~\cite{refBarlow}. The significance of the deviation, $\zeta$, is defined as $\zeta=\frac{|\mathcal{B}_{\mathrm{nominal}}-\mathcal{B}_{\mathrm{test}}|}{\sqrt{|\sigma_{\mathcal{B}_{\mathrm{nominal}}}^{2}-\sigma_{\mathcal{B}_{\mathrm{test}}}^{2}|}}$, where $\mathcal{B}_{\mathrm{nominal}}$ denotes the nominal BF, $\mathcal{B}_{\mathrm{test}}$ denotes the BF obtained under the test condition, and $\sigma_{\mathcal{B}_{\mathrm{nominal}}}$ and $\sigma_{\mathcal{B}_{\mathrm{test}}}$ are their corresponding statistical uncertainties. If $\zeta<2$ is satisfied for all test conditions, the difference in measured BFs is attributed to statistical fluctuations, rendering the uncertainty negligible. Otherwise, the largest difference corresponding to the maximum value of $\zeta$ is taken as the uncertainty. To obtain the $\zeta$ distribution, we vary the condition $\chi_{\mathrm{4C}}^{2}-\chi_{\mathrm{4C},1\gamma}^{2}<0$ by $\pm5$ in intervals of 1 to evaluate the BFs. Since all the resulting values of $\zeta$ for the adjusted conditions are less than 2, the uncertainty is deemed negligible.

The uncertainties related to the mass window veto criteria are estimated by expanding and shrinking the corresponding windows by the fitted mass resolution. The larger differences in the measured BFs are taken as the uncertainties.

The uncertainties related to the ST fit model are estimated as follows. The uncertainty due to the fit range is estimated by varying the fit range by $\pm2$ times the fitted mass resolution, and the largest difference in the measured BF is taken as the uncertainty. The uncertainty from the signal shape is estimated by varying the mass and width of the Breit-Wigner function by $\pm1\sigma$. Additionally, an uncertainty arises from the mass resolution difference between data and MC simulation, estimated by convolving the original signal function with a Gaussian function. The largest difference in the signal yield is considered as the uncertainty. The uncertainty associated with the background shape is determined by replacing the nominal background function with an alternative function, which consists of the inclusive MC shape and a second-order Chebyshev polynomial function. The difference in the signal yield is taken as the uncertainty due to the background shape.

The uncertainties related to the DT fit model are estimated as follows. The uncertainty due to the fit range is assessed in the same manner as for the ST fit model. The uncertainty from the signal shape is estimated by changing the MC signal shape. Specifically, we vary the mass and width of $\Xi(1530)^{-}$ by $\pm1\sigma$ when generating exclusive signal MC samples. The largest difference in the signal yield is taken as the uncertainty due to the signal shape. The uncertainty due to the background shape is estimated using 1000 groups of toy MC samples.
In the nominal fit, the background functions are fixed to the MC-derived shapes.
We release the constraints on each background function individually and perform fits to toy MC samples generated with the nominal fit function.
The number of signal events from the 1000 groups of toy MC samples follows a Gaussian distribution. The largest difference between the mean value of the Gaussian distribution and the nominal value of the signal yield is taken as the uncertainty due to the background shape.

The uncertainties for the BFs used in the calculations, such as $\mathcal{B}(\bar{\Xi}^{+}\to\bar{\Lambda}\pi^{+})$ and $\mathcal{B}(\bar{\Lambda}\to\bar{p}\pi^{+})$, are taken from the PDG~\cite{refPDG}. The uncertainty due to the MC statistics is calculated as $\sqrt{\frac{1-\epsilon}{N_{\mathrm{tot}}\cdot\epsilon}}$, where $N_{\mathrm{tot}}$ is the total number of generated MC events. The uncertainty of the total number of $\psi(3686)$ events in data is 0.5\%~\cite{refpsipnumber}.

In addition, there is an uncertainty associated with the effect of the interference between the resonance decay and continuum process in the measurement of the BF of $\psi(3686)\to\bar{\Xi}^{+}\Xi(1530)^{-}$. To estimate the interference effect, we use the method from Ref.~\cite{refinterference_err}. The maximum impact of the interference term with respect to the resonance term is defined as $r_{R}^{\mathrm{max}}$,
\begin{equation}\begin{split}\label{eq:sys_interference}
r_{R}^{\mathrm{max}}=\frac{2}{\hbar c}AB,\ A=\sqrt{\frac{\sigma_{c}}{\mathcal{B}_{f}}},
\end{split}\end{equation}
where $\mathcal{B}_{f}$ is the BF of $\psi(3686)\to\bar{\Xi}^{+}\Xi(1530)^{-}$ obtained from our measurement, the factor $B=6.74\ \mathrm{GeV}/c^{2}$ is a constant depending on the resonance parameters quoted from Ref.~\cite{refinterference_err}, and $\hbar$ is the reduced Planck constant. The $r_{R}^{\mathrm{max}}$, 6.6\%, is taken as the uncertainty of interference.

\begin{table*}[htbp]
\centering
\caption{Relative systematic uncertainties (\%) in the BF measurements. The symbol ``$\ldots$" indicates the absence of systematic uncertainty from the corresponding source.}
\begin{tabular}{cccc}
\hline\hline
Source & $\psi(3686)\to\bar{\Xi}^{+}\Xi(1530)^{-}$ & $\Xi(1530)^{-} \to \Xi^{0}\pi^{-}$ & $\Xi(1530)^{-} \to \Xi^{-}\pi^{0}$ \\ \hline
$\bar{\Xi}^{+}$ reconstruction & 1.3 & $\ldots$ & $\ldots$ \\
$\Lambda$ reconstruction & $\ldots$ & 1.0 & 1.0 \\
$\pi^{-}$ tracking and PID & $\ldots$ & 1.4 & 1.4 \\
$\pi^{0}$ reconstruction & $\ldots$ & 2.0 & 2.0 \\
$\bar{\Lambda}$ mass window & negligible & $\ldots$ & $\ldots$ \\
$\Lambda$ mass window & $\ldots$ & 0.1 & 0.1 \\
$\bar{\Xi}^{+}$ mass window & negligible & $\ldots$ & $\ldots$ \\
4C kinematic fit & $\ldots$ & negligible & 0.1 \\
$\chi^{2}_{\mathrm{4C}}<\chi^{2}_{\mathrm{4C},1\gamma}$ & $\ldots$ & negligible & negligible \\
$|M_{\pi^{+}\pi^{-}}^{\mathrm{recoil}}-m_{J/\psi}|>7.5\ \mathrm{MeV}/c^{2}$ & 0.3 & $\ldots$ & $\ldots$ \\
$|M_{\pi^{+}\pi^{-}\pi^{0}}^{\mathrm{recoil}}-m_{J/\psi}|>15\ \mathrm{MeV}/c^{2}$ & $\ldots$ & 0.6 & 0.4 \\
$|M_{\pi^{0}\pi^{0}}^{\mathrm{recoil}}-m_{h_{c}}|>10\ \mathrm{MeV}/c^{2}$ & $\ldots$ & 0.8 & 0.8 \\
ST fit model & 6.5 & 6.5 & 6.5 \\
DT fit model & $\ldots$ & 2.1 & 2.3 \\
BFs & 0.8 & 0.8 & 0.8 \\
MC statistics & 0.2 & 0.6 & 0.6 \\
$N_{\psi(3686)}$ & 0.5 & $\ldots$ & $\ldots$ \\ \hline
Total & 6.7 & 7.5 & 7.5 \\
Interference & 6.6 & $\ldots$ & $\ldots$ \\
\hline\hline
\end{tabular}
\label{table:sysuncertainty_summary}
\end{table*}

\section{Summary}
Using $(2712.4\pm14.3)\times10^{6}$ $\psi(3686)$ events collected by the BESIII detector,
we report the first measurement of the absolute BFs of the decays $\Xi(1530)^{-}\to\Xi^{0}\pi^{-}$ and $\Xi(1530)^{-}\to\Xi^{-}\pi^{0}$.
The two BFs are measured assuming full correlation based on isospin symmetry, yielding $\mathcal{B}(\Xi(1530)^{-} \to \Xi^{0}\pi^{-})=(61.4\pm4.5\pm4.6)\%$ and $\mathcal{B}(\Xi(1530)^{-} \to \Xi^{-}\pi^{0})=(29.7\pm2.2\pm2.2)\%$. The combined BF of the two decays is $\mathcal{B}(\Xi(1530)^{-} \to (\Xi\pi)^{-})=(91.1\pm6.7\pm6.8)\%$, where the common systematic uncertainties in the two decay channels are only considered once. Here, the first uncertainties are statistical and the second are systematic. This result is consistent with the PDG value of 100\%~\cite{refPDG}. Additionally, we update the BF of $\psi(3686)\to\bar{\Xi}^{+}\Xi(1530)^{-}$, yielding $\mathcal{B}(\psi(3686)\to\bar{\Xi}^{+}\Xi(1530)^{-})=(8.67\pm 0.52\pm0.58\pm0.57)\times 10^{-6}$, where the first uncertainty is statistical, the second is systematic related to event selection and the fit model, and the third is systematic associated with the interference effect. This result is consistent with the previous measurement but with much higher precision~\cite{refpsipToXiXi1530}. The measurement of $\psi(3686)\to\bar{\Xi}^{+}\Xi(1530)^{-}$ indicates a violation of the SU(3)-flavor symmetry in this decay, thereby reinforcing the generality of SU(3)-flavor symmetry breaking. Our work contributes to the study of QCD and the quark model, and provides additional insights for the investigation of the $\Xi$ baryon spectroscopy.

\acknowledgements


The BESIII Collaboration thanks the staff of BEPCII (https://cstr.cn/31109.02.BEPC) and the IHEP computing center for their strong support. This work is supported in part by National Key R\&D Program of China under Contracts Nos. 2025YFA1613900, 2023YFA1606000, 2023YFA1606704; National Natural Science Foundation of China (NSFC) under Contracts Nos. 11635010, 11935015, 11935016, 11935018, 12025502, 12035009, 12035013, 12061131003, 12192260, 12192261, 12192262, 12192263, 12192264, 12192265, 12221005, 12225509, 12235017, 12342502, 12361141819; the Chinese Academy of Sciences (CAS) Large-Scale Scientific Facility Program; the Strategic Priority Research Program of Chinese Academy of Sciences under Contract No. XDA0480600; CAS under Contract No. YSBR-101; Shanghai Leading Talent Program of Eastern Talent Plan under Contract No. JLH5913002; 100 Talents Program of CAS; The Institute of Nuclear and Particle Physics (INPAC) and Shanghai Key Laboratory for Particle Physics and Cosmology; ERC under Contract No. 758462; German Research Foundation DFG under Contract No. FOR5327; Istituto Nazionale di Fisica Nucleare, Italy; Knut and Alice Wallenberg Foundation under Contracts Nos. 2021.0174, 2021.0299, 2023.0315; Ministry of Development of Turkey under Contract No. DPT2006K-120470; National Research Foundation of Korea under Contract No. NRF-2022R1A2C1092335; National Science and Technology fund of Mongolia; Polish National Science Centre under Contract No. 2024/53/B/ST2/00975; STFC (United Kingdom); Swedish Research Council under Contract No. 2019.04595; U. S. Department of Energy under Contract No. DE-FG02-05ER41374.

\end{document}